\documentclass[twocolumn,eqsecnum,aps,prd,
amsmath,amssymb,nofootinbib,floatfix]{revtex4}

\usepackage{bm}
\usepackage{graphicx,epsf}

\newcommand\lsim{\lesssim}

\newcommand\gsim{\gtrsim}

\renewcommand\({\left(}

\renewcommand\){\right)}

\renewcommand\[{\left[}

\renewcommand\]{\right]}

\newcommand\eq[1]{Eq.~(\ref{#1})}

\newcommand\eqreff[1]{(\ref{#1})}

\newcommand\ee{\end{equation}}

\newcommand\be{\begin{equation}}

\newcommand\eea{\end{eqnarray}}

\newcommand\bea{\begin{eqnarray}}

\newcommand\mpl{M_{\rm P}}

\def\calp{{\cal P}}

\newcommand\GeV{\,\mbox{GeV}}

\newcommand\quarter{^{1/4}}

\begin{document}

\title{Inflation models after WMAP year three}

\author{Laila Alabidi and David H.\ Lyth}

\affiliation{Physics Department, Lancaster University,LA1 4YB}


    \begin{abstract}
The survey of models in astro-ph/0510441 is updated. 
For the first time, a large fraction
of the models is ruled out at more than $3\sigma$.
\end{abstract}

    \maketitle

        \section{Introduction}\label{intro}

In a recent paper \cite{al} we discussed a range of
models for the origin of the 
curvature perturbation and the tensor perturbation,  including
constraints on the spectral index $n$ coming from WMAP year one data. 
In this note we update the discussion to include WMAP year three data
\cite{wmap3}.
The   models assume  that the curvature perturbation is generated
from the vacuum fluctuation of the inflaton field, so that it is 
directly related to the inflationary potential.\footnote
{Models where instead the curvature perturbation is generated from
the vacuum fluctuation of some curvaton-type field, and
 their status after WMAP year three,  are  considered 
elsewhere \cite{p052}.}
 Some of them
 work with the type of field theory that is 
usually invoked when considering extensions of the Standard Model,
while  others work within a framework  derived
more or less directly from string theory.

The prediction for $n$ typically depends on $N$, the number of 
$e$-folds of slow-roll inflation occurring after the observable Universe
leaves the horizon. With a high inflation scale,  and  radiation and/or
matter domination between the end of inflation and nucleosynthesis,
\be
N= 54\pm 7
\label{smallrange}
.
\ee
More generally the range has to be
\be
14 < N < 75
\label{bignrange}
,
\ee 
the lower bound coming from the requirement to form early objects weighing
 a million solar masses, and the upper bound from imposing $P/\rho< 1$ on the 
pressure and energy density \cite{andrewn}. 

Following \cite{al}, we call
a model small-field if the change $\Delta\phi_N$ of  the inflaton field
during the $N$ $e$-folds is $\lsim \mpl$, and large-field if it is
$\gg \mpl$.

\section{Small-field models}

\subsection{A class of allowed models}

Assuming that the  tensor fraction  $r$
is negligible and that the spectral tilt $n-1$ is practically 
constant while cosmological scales leave the horizon,
 the WMAP constraint is 
\be
n= 0.948^{+0.015}_{-0.018}
.
\ee
This is actually the  constraint obtained by combining WMAP data with
the SDSS galaxy survey,  but it hardly changes if some other
data sets are used, including WMAP alone. 

Essentially the same constraint was
obtained using WMAP year one data, in conjunction with the  2dF galaxy 
survey alone \cite{2dfwmap} or with 2dF and other data sets
 \cite{wbs2}, but a higher result compatible with 
$n=1$ was obtained  using WMAP year one data alone or WMAP with \cite{sdsswmap}
SDSS.
The crucial point now is that even in the last two cases
{\em the scale-invariant value $n=1$ is excluded at around the $3\sigma$
level}.

For small-field models with a concave-downward potential, $\epsilon\lsim
0.0002$ \cite{hilltop}. Then 
 the prediction \cite{ll92} 
 $n=1+2\eta-6\epsilon$ becomes just\footnote
{We adopt the definitions \cite{ll92} $2\epsilon=(\mpl V'/V)^2$ and
$\eta = \mpl^2 V''/V$, with $V(\phi)$ the inflationary potential.}
$n = 1+ 2\eta$.
For a  wide class of concave-downward models this becomes \cite{treview}
\be
n = 1-\frac{p-1}{p-2} \frac2{N}
\label{nnew}
,
\ee
with $p\gsim 3$ or $p\leq 0$. (Here $N$ is actually $N(k)\equiv N-\ln(H_0/k)$,
but the variation presumably 
is negligible over the range $\Delta N\sim 10$ over which
the observational constraint applies.)
For these models, the observed normalization
of the spectrum requires a high inflation scale, so that \eq{smallrange}
will be appropriate for a standard post-inflationary cosmology, but still
the tensor fraction is negligible.

The case $p\leq 0$ is realised in some
hybrid inflation models in which the potential necessarily steepens 
significantly towards the end of inflation. 
The case $p< 0$ 
corresponds to a potential
 \be
            V \simeq V_0 \[ 1- \( \frac\phi\mu \)^p \]
\label{vnew}
                 ,
  \ee
with $V_0$ dominating so as to permit inflation.
This can come from 
mutated hybrid inflation 
\cite{mutated,inverted}, with integer values of $p$
favoured but not mandatory. With integral $p$
it can  also come from  $N=2$ supergravity
\cite{juann2} or  D-brane cosmology  \cite{dbraneinf}.
The  limit $p\to 0$ corresponds to a logarithmic potential achieved in
 the simplest 
and perhaps unrealistic version of $F$-term 
\cite{cllsw,fterm,ewanexp}  or $D$-term
\cite{ewanexp,dterm} hybrid inflation. The limit
$p\to -\infty$ corresponds to an exponential potential, which may be 
generated by  a kinetic term passing through zero \cite{ewanexp} or by
 by appropriate  non-Einstein gravity (non-hybrid)
inflation \cite{treview} (see also \cite{cq}).

The  case $p\gsim 3$ also corresponds to \eq{vnew}. 
This case is attractive because it gives the 
 potential  a maximum, at which
 eternal inflation can take  place  providing the 
initial condition for the subsequent slow roll
\cite{eternal1,eternal2,hilltop}.
As \eq{vnew} 
 is only supposed to be an approximation lasting for a sufficient number
of $e$-folds, $p$  need not be  an integer, but it has to be well above
$2$ for the  prediction \eqreff{nnew} to hold.
It could correspond to non-hybrid inflation with  $\Delta\phi \ll\mpl$
(New Inflation \cite{new,sv})
 or else with $\Delta\phi\sim \mpl$
(Modular Inflation which has a long history 
\cite{andrei83,modular1,graham}
and is currently under
 intense investigation in the context of string theory
\cite{modular2,more,west}). 
It could also correspond to mutated hybrid inflation \cite{inverted}, 
or else 
\cite{hilltop} to 
one of the $p\leq 0$ models,  modified by the addition of 
a non-renormalizable term. 

An  attractive proposal which can
give \eq{nnew} is to make the inflaton a pseudo-Nambu-Goldstone boson
so that the flatness of its potential is protected by a symmetry.
Realizations of this proposal  include a two-component model
giving $p=3$ \cite{cs} and a hybrid model giving $p=0$ \cite{pseudonat}.
(A different proposal for ensuring the flatness is described in
\cite{glm} based on earlier work \cite{cllsw,ewanexp}, but it has not been
carried through to the point where a definite form for the potential is
proposed.)

\begin{figure}
 \centering\includegraphics[angle=270, width=1.0\columnwidth,totalheight=2.5in]{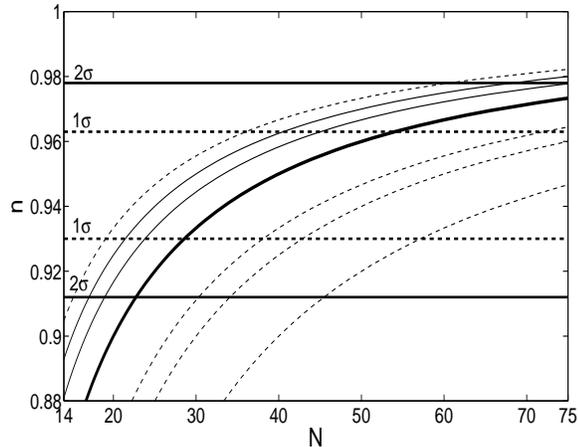}
               \caption{The prediction  \eqreff{nnew} for different $p$.
                The bold full line is the limit $|p|\to\infty$. Above it from top down are the
                lines $p=0$, $-2$  and $-4$, and below it from bottom up are the lines
                $p=3$, $4$ and $5$. The observational
bounds from \cite{wmap3} are indicated.}
               \label{nvsN}
            \end{figure}

\begin{figure}
                \centering\includegraphics[angle=270,width=1.0\columnwidth,totalheight=2.5in]{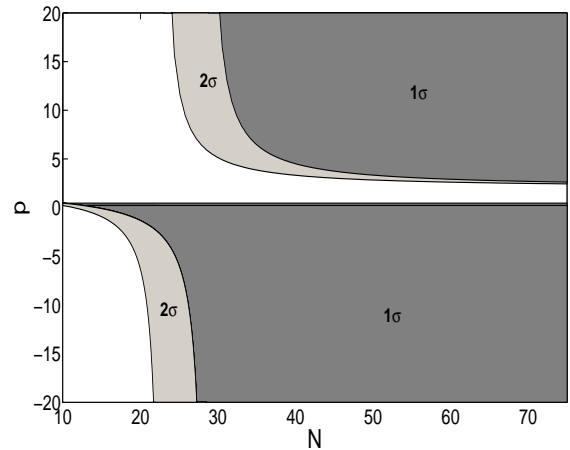}
                 \caption{The regions excluded by the observational bounds
from \cite{wmap3}, for the parameter $p$ in the prediction \eqreff{nnew}.}
                 \label{pvsN}
            \end{figure}

\begin{figure}
                \centering\includegraphics[angle=270,
                width=1.0\columnwidth, totalheight=2in]{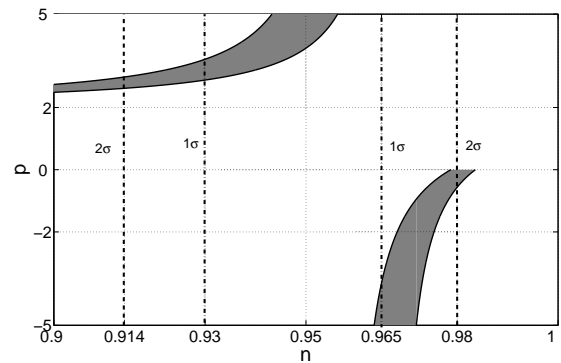}
                \caption{The prediction \eqreff{nnew}
for $N=54\pm 7$.}
                \label{pvsn}
            \end{figure}

In Figure \ref{nvsN}, the  prediction \eqref{nnew}
is shown against $N$ for a few values of $p$. 
Very low values of $N$ are forbidden, as 
 is seen clearly in  Figure  \ref{pvsN}.
With $N$ in the reasonable range \eqreff{smallrange}, the 
prediction $n(p)$  is shown in Figure \ref{pvsn}, where we see that all
values of $p$ are just about allowed at the $2\sigma$ level. 

        \begin{figure}
        \centering\includegraphics[angle=270,width=1.0\columnwidth]{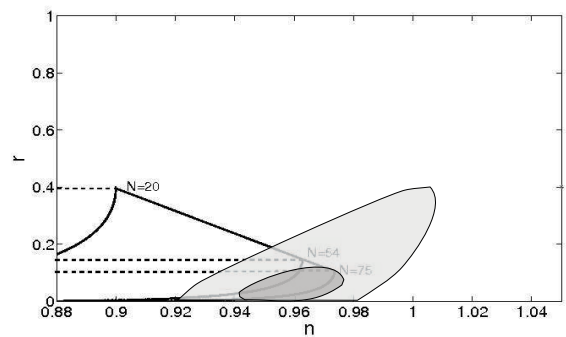}
        \caption{The curved lines are the Natural Inflation predictions for 
$N=20$, $N=54$ and $N=75$, 
and the horizontal lines are the corresponding multi-component
        Chaotic Inflation predictions. 
The junction of each pair of lines corresponds
        to single-component Chaotic Inflation.  
The regions allowed by observation
        with various assumptions are taken from \cite{wmap3}.}
                                    \label{rvsn}
                                \end{figure}

These cases give a spectral index more or less in agreement with the observed
 one because their prediction is $n-1\simeq -1/N$.
The case of \eq{vnew} with $p=2$
 is quite different. The tilt is now 
$n-1=-2\mu^2\mpl^2/V$, which might have had any value in the range
$-1\ll n-1 <0$. It depends on the parameters of the potential,
not just on its functional form as in the previous case.

Now that observation requires such a small tilt, the  
 case $p=2$ actually looks rather problematic because it requires a rather
abrupt steepening  of the potential after cosmological scales leave the 
horizon. This is difficult to achieve in a non-hybrid model;
for instance the Little-Higgs proposal of \cite{pseudonat}
seems not to give a sufficiently abrupt end. It can be achieved in
an  inverted hybrid model \cite{inverted} by choice of parameters, but this 
typically involves fine-tuning \cite{steveinvert}, and the  
negative coupling of the inflaton to the waterfall field is non-standard
and difficult to achieve in the context of supersymmetry.

\subsection{Models ruled out}

 If the observation of negative tilt
 holds up it will represent a very significant development.
Speaking generally, it avoids the criticism that
 $n=1$ might have had some simple
explanation, overlooked so far, which has nothing to do with field theory
or inflation. 
Within the context of slow-roll inflation, $n=1$ excludes several possibilities
for the inflationary potential in a small-field model.

\paragraph{Concave-upward potentials}
A   concave-upward  potential give positive tilt if $2\eta>6\epsilon$.
That is generally the case for small-field models. In particular it is true
for small-field models with 
\be
V= V_0 \[1 + \( \frac\phi\mu \)^p \]
\label{cup}
.
\ee 
Indeed, $V_0$ must  dominate  to achieve small-field inflation, but 
then $\epsilon \sim (\phi/\mpl)^2 \eta^2\ll \eta$.
 An  attractive realization of this potential is
the original hybrid model with $p=2$\label{originalhybrid}.
An integer $p\geq 3$  also corresponds to tree-level hybrid inflation,
\cite{andreihybrid,hybrid2,inverted} while the case  $p\leq -1$
corresponds to dynamical supersymmetry breaking \cite{dsb}. 
These are  less attractive because small-field inflation occurs only 
over a limited range of $\phi$ and 
 it is not clear how the field is supposed to arrive within this range
\cite{treview}.

\paragraph{Very flat potentials}
If the  potential is very flat, $\epsilon$ and $\eta$ will be negligible and
so will the tilt $n-1$. This happens in some
 models which seek to explain the inflationary
scale   $V$ by identifying it with the scale of supersymmetry breaking in the 
vacuum \cite{supernatural,steve,dr}. The models make the very potential flat
in order to reproduce the observed spectrum  
$\calp_\zeta = (5\times 10^{-5})^2$ with the formula
 $\calp_\zeta  \simeq  (V/\epsilon \mpl^4)$ and  the low scale
$V\quarter\lsim 10^{10}\GeV$ assumed for supersymmetry breaking. 
Such models  are accordingly ruled out
by the observed tilt. 

\paragraph{Running Mass inflation} 
The idea of running mass inflation is to use a 
loop correction,  to flatten a tree-level  potential, which would otherwise
be too steep for inflation.
The model of
\cite{ewanrunning,cl} uses  a tree-level quadratic
potential (\eq{cup} with $p=2$) modified by a 
one-loop correction. It is assumed that the tree-level potential
has $\eta\sim 1$, which is the generic supergravity value and marginally
spoils inflation.
 This model gives significant positive running
 of the spectral index $dn(k)/dk>0$, which  was  allowed by 
the  WMAP year one  data \cite{ourrunning}
 if $n$  passed through zero around the the middle of  the cosmological
range of scales. The model
  presumably is ruled out by
the  WMAP  three year data, which allows  $n(k)$ to pass through 
1  only in the negative direction. The  alternative model of 
\cite{ks} makes inflaton is a two-component modulus. It typically gives
either negligible tilt or tilt with rather strong running, but 
further investigation is needed to see whether it is ruled out.

\paragraph{Generic   modular inflation and supergravity}
A string theory modulus is expected generically to  
have a potential of the form
$V=V_0f(\phi/\mpl)$, with $f(x)$  and its low derivatives  of order 1
at a generic point in the range $\phi\lsim \mpl$. Near a maximum this gives
$\eta\sim -1$ which only marginally allows inflation and gives
$n-1\sim -1$. A similar result,  $|\eta|\gsim 1$, is expected in a generic
supergravity theory for any field.
One of the most significant consequences of the  bound
$|n-1|\lsim 0.1$, which observation has provided 
in recent years, is that $|\eta|$  has to be reduced
below its generic value  by more than a factor  $10$. The new result
for $n-1$ confirms that, but it also assures us that we will not have
to go much further.
Inflation based on a modulus
and/or supergravity requires  fine-tuning of $\eta$ at the few percent level,
but not worse. More fine-tuning is typically needed though, to 
 stabilize fields other than $\phi$.

\section{Large-field models}

Large-field models allow an observable tensor fraction $r$. The
single-component models are chaotic inflation \cite{chaotic},
the  multi-component version of that \cite{starob85,Nflation}, and 
Natural Inflation \cite{natural}.
The  situation for these models  is illustrated in Figure
\ref{rvsn}. (It shows the  WMAP/SDSS constraint, but
  but it hardly changes if WMAP is combined with 
other data sets.)
There is no dramatic change from the situation with earlier 
constraints derived from WMAP year-one. 

The generic prediction for a chaotic inflation potential
 $V\propto \phi^\alpha $ is
\bea
r&=& \frac{4\alpha}{N} \\
n-1 &=& 2\eta- 6\epsilon = - \frac{2+\alpha}{2N}
.
\eea
As pointed out already \cite{wmap3}, the year-three WMAP data rule out rather
firmly $\alpha\geq 4$. Interestingly enough, the allowed case $\alpha=2$
is also the best-motivated  one in the context of received ideas about
field theory  \cite{extranat,pseudonat,Kim:2004rp}, 
because it is reproduced by a Natural Inflation potential with a large period.

\section{Summary and outlook}

Over the last twenty-five years many  field-theory
models of slow-roll inflation have been proposed. We have seen that the
 WMAP year three results
for $n$ and $r$  rule out a large fraction of these models.
The remainder 
seem to be in three broad classes; large field models,
small-field models giving the prediction \eqreff{smallrange} with $p\leq 0$,
and small-field models giving the same prediction with $p\gsim 3$.
Within a few years, the
 PLANCK  result for $n$ and the Clover result for
$r$ will almost certainly rule out at least two of these 
classes, and provide some discrimination within the remaining class.

            \begin{acknowledgments}
 DHL
is supported by
     PPARC grants PPA/G/S/2002/00098, PPA/G/O/2002/00469, PPA/Y/S/2002/00272,
            PPA/V/S/2003/00104  and EU grant MRTN-CT-2004-503369.
            LA thanks Lancaster University for the award of the studentship from Dowager Countess Eleanor Peel
            Trust.
\end{acknowledgments}

\end{document}